\documentstyle[12pt]{article}
\begin{document}
\begin{center}
{\Large\bf On generalized fractional kinetic equations}\\[1cm]
{\large R.K. Saxena}\\
Department of Mathematics and Statistics, 
Jai Narain Vyas University Jodhpur
342001, INDIA\\[0.5cm]
{\large A.M. Mathai}\\ 
Department of Mathematics and Statistics, McGill University,\\
805 Sherbooke Street West, Montreal, CANADA H3A 2K6\\[0.5cm]
{\large H.J. Haubold}\\ 
Office for Outer Space Affairs, United Nations,\\ 
P.O. Box 500, A-1400 Vienna, AUSTRIA\\
\end{center}
\noindent                            
{\bf Abstract}\par
In a recent paper,   Saxena et al.[1] developed the solutions of three generalized fractional kinetic equations in terms of the Mittag-Leffler functions. The object of the present paper is to further derive the  solution of further generalized fractional kinetic equations. The results are obtained in a compact form in terms of generalized Mittag-Leffler functions. Their relation to fundamental laws of physics is briefly discussed.
\section{Introduction}\par

The fundamental laws of physics are written as equations for the time evolution of a quantity $X(t), dX(t)/dt=-AX$, where this could be Maxwell's equations or Schr\"odinger's equation (if A is limited to linear operators), or it could be Newton's law of motion or Einstein's equations for geodesics (if A may also be a nonlinear operator [2,3,27]). The mathematical solution (for linear operators A) is $X(t) = X(0)exp\{-At\}$.

In thermodynamical or statistical applications one is mostly interested in mean values of the quantity $<X(t)>$. In this case, A is a characteristic time scale $A^{-1}=\tau$
in the evolution equation for $<X(t)>$. It then follows that $<X(t)>$ decays exponentially toward equilibrium $<X(t)> = <(X(0)>exp\{-t/\tau\}.$

In 1988, Tsallis [4] generalized the entropic function of Boltzmann-Gibbs statistical mechanics, $s=-\int dx p(x)lnp(x),$
to nonextensive statistical mechanics with $S_q[p]=\left\{1-\int dx[p(x)]^q\right\}/(q-1)$
that leads to q-exponential distributions $p_q(x)\propto\left[1-(1-q)x^2/kT\right]^{1/(1-q)}.$
Such a distribution reduces to Gaussian distribution for $q = 1$ and for $q = 2$ to a Cauchy-Lorentz distribution, to name two examples. In an attempt to incorporate L\'evy distribution into statistical mechanics, Tsallis has also shown that the above distribution becomes a L\'evy distribution for $q > 5/3$. Recently, Tsallis [4] used the mathematical simplicity of reaction-type equations, $dX/dY = Y^q$, to emphasize the natural outcome of the above distribution function $p_q(x)$ which corresponds exactly to the solution of the reaction equation of nonlinear type. The solution has power-law behavior. In the following we show that the fractional generalization of the linear reaction-type equation also leads to power-law behavior. In both cases, solutions can be expressed in terms of generalized Mittag-Leffler functions. 
  
\section{Generalized Mittag-Leffler function}
 A generalization of  the Mittag-Leffler function [5,6]
\begin{equation} 
E_\alpha(z):=\sum^\infty_{n=0}\frac{z^n}{\Gamma(\alpha n+1)},
\end{equation}                                                                
and its generalized form
\begin{equation}
E_{\alpha,\beta}(z):=
\sum^\infty_{n=0}\frac{z^n}{\Gamma(\alpha n+\beta)}\;\;\;(\alpha, \beta \in C, Re(\alpha)>0)
\end{equation}      
was introduced by Prabhakar [7] in terms of the series representation
\begin{equation}
E^\gamma_{\alpha, \beta}(z):=\sum^\infty_{n=0}\frac{(\gamma)_nz^n}{\Gamma(\alpha n+\beta)(n!)}, \;\;\;(\alpha, \beta, \gamma \in C, Re(\alpha)>0),
\end{equation}               
where $(\gamma)_n$  is  Pochammer's symbol defined by
\begin{equation}
(\gamma)_n=\left\{^{1,n=0}_{\gamma(\gamma+1)\ldots(\gamma+n-1), n\in N}
\right.\;\;  
\gamma \neq 0                                                                  
\end{equation}
It is an entire function of order $\rho = [Re(\alpha)]^{-1}$  [7].  This function  has been studied  by Wiman [8,9], Agarwal [10], Humbert [11]  and Humbert and Agarwal [12] and several others. Some special cases of (3) are given below:
\begin{equation}
(i)\;\;\; E_\alpha(z)=E^1_{\alpha, 1}(z),
\end{equation}
\begin{equation}
(ii)\;\; E_{\alpha, \beta}(z)= E^1_{\alpha, \beta}(z),
\end{equation}
\begin{equation}
(iii)\; \Phi(\beta, \gamma; z)=\; _1F_1(\beta;\gamma;z) = \Gamma(\gamma)E_{1,\gamma}^\beta(z),
\end{equation}                                                              
where $\Phi(\beta, \gamma;z)$    is  Kummer's confluent hypergeometric function defined in 
  Erd\'elyi et al. ([13], p. 248, eq.1]).
        Mellin-Barnes integral representation for the  function defined by (3)  follows from the integral 
\begin{equation}
E^\gamma_{\alpha,\beta}(z)=\frac{1}{2\pi\omega\Gamma(\gamma)}\int_\Omega \frac{\Gamma(-s)\Gamma(\gamma+s)(-z)^s}{\Gamma(\beta+s\alpha)}ds,
\end{equation}
where $\omega=(-1)^{1/2}$. The contour $\Omega$ is a straight line parallel to the imaginary axis separating the poles of $\Gamma(-s)$ at the points 
$s = \nu\;\;(\nu=0,1,2,\ldots)$ from  those of $\Gamma(\gamma+s)$  at the points  $s = -\gamma-\nu\;\;(\nu=0,1,2,\ldots)$.
The poles of the integrand of (8) are assumed to be simple. (8) can be established by calculating the residues at the poles of $\Gamma(-s)$  at   the points , $s =\nu\;\;(\nu=0,1,2,\ldots,).$
 It follows from (8) that $E^\gamma_{\alpha,\beta}(z)$ can be represented in the form 
\begin{equation}
E_{\alpha,\beta}^\gamma(z)=\frac{1}{\Gamma(\gamma)}H_{1,2}^{1,1}\left[-z|^{(1-\gamma,1)}_{(0,1),(1-\beta,\alpha)}\right], (Re(\alpha)>0;\alpha,\beta,\gamma \in C),
\end{equation}
where $H^{1,1}_{1,2}(z)$ is the H-function. A detailed account of the theory and applications of the H-function is available from   Mathai and Saxena [14]. This function can also be represented by 
\begin{equation}
E^\gamma_{\alpha,\beta}(z)=\frac{1}{\Gamma(\gamma)}\;_1\Psi_1\left[^{(\gamma,1)}_{\beta,\alpha)};z\right],
\end{equation}                                                                    
where $_1\Psi_1(z)$   is a special case of  Wright's generalized hypergeometric function $_p\Psi_q(z)$
      [15,16] ; also see,  Erd\'elyi  et al. ([13], Section 4.1 ), defined by 

\begin{equation}
_p\Psi_q\left[^{(a_1,A_1),\ldots,(a_p,A_p)}_{(b_1,B_1),\ldots,(b_q,B_q)};z\right]=\sum^\infty_{n=0}\frac{\prod^p_{j=1}\Gamma(a_j+A_jn)}{\prod ^q_{j=1}\Gamma(b_j+B_jn)}\frac{z^n}{(n)!},
\end{equation}
where  $1+\sum_{j=1}^q B_j-\sum^p_{j=1}A_j\geq 0$ (equality only holds for appropriately bounded $z$).
        When $\gamma=1$, (9) and (10) give rise to  (12) and (13) given below:
\begin{eqnarray}
E_{\alpha,\beta} (z) & = & _1\Psi_1\left[^{(1,1)}_{(\beta,\alpha)};z\right],\\
& = & H^{1,1}_{1,2}\left[-z\left[^{(0,1)}_{(0,1),(1-\beta, \alpha)}\right]\right.,
\end{eqnarray}
where $Re(\alpha)>0; \alpha, \beta, \gamma \in C.$

If we further take $\beta=1$ in (12) and (13) we find that
\begin{eqnarray}
E_\alpha(z) & = &_1\Psi_1\left[^{(1,1)}_{(1,\alpha)};z\right],\\
& = & H^{1,1}_{1,2}\left[-z|^{(0,1)}_{(0,1),(0,\alpha)}\right],
\end{eqnarray}                                                                               
for $Re(\alpha)>0, z\in C.$  
The following integral gives the Laplace transform of $E^\gamma_{\alpha,\beta}(z)$.
\begin{equation}
\int^\infty_0 e^{-pt}t^{\beta-1}E^{\gamma}_{\alpha,\beta}(at^\alpha) dt=p^{-\beta}(1-ap^{-\alpha})^{-\gamma},
\end{equation}      
where $Re(p) > |a|^{\frac{1}{Re(\alpha)}}, Re(\beta)>0, Re(p)>0,$
which can be established by means of the Laplace integral
\begin{equation}
\int^\infty_0 e^{-pt}t^{\rho-1}dt = \frac{\Gamma(\rho)}{p^\rho}, Re(p)>0, Re(\rho)>0.
\end{equation}   
For $\gamma=1,$ (16) reduces to an elegant formula
\begin{equation} 
        \int^\infty_0 e^{-pt}t^{\beta-1} E_{\alpha, \beta}(at^\alpha)dt = p^{-\beta}(1-ap^{-\alpha})^{-1},
\end{equation}
where $Re(\beta) > 0, Re(p) > 0 , |p| > |a|^{\frac{1}{Re(\alpha)}}$ .
        In an attempt to investigate the functions which when fractionally differentiated (of any order) reappear,  Hartley and Lorenzo [17]  came across a  special function of  the form 
\begin{eqnarray}
F_q[-a,t] &=& t^{q-1}\sum^\infty_{n=0}\frac{(-a)^nt^{nq}}{\Gamma(nq+q)}, Re(q)>0,\\
& = & t^{q-1} E_{q,q}(-at^q).
\end{eqnarray}             
This function has been studied  earlier by  Robotnov [18,19] in connection with hereditary integrals for application to continuum mechanics. The Laplace transform of this function is given by 
\begin{equation}
L[F_q(a;t)] = \frac{1}{p^q-a}, Re(q)>0.
\end{equation}

A generalization of the F-function is presented by Lorenzo and Hartley [20] by means of the  following series representation:
\begin{eqnarray}
R_{\nu,\mu}[a,c,t]&=& \sum^\infty_{n=0}\frac{a^n(t-c)^{(n+1)\nu-\mu-1}}{\Gamma[(n+1)\nu-\mu]}, t>c>0\\
& = & (t-c)^{\nu-\mu-1} E_{\nu,\nu-\mu}[a(t-c)^\nu], t>c>0.
\end{eqnarray}                                
The Laplace transform of the R-function is derived  by  Lorenzo and Hartley [20] in the form
 \begin{equation}
L\left\{R_{\nu,\mu}(a,c,t)\right\}= \frac{e^{-cp}p^\mu}{p^\nu-a}, \;\;c\geq 0,
\end{equation}
where $Re(\nu-\mu)>0, Re(p)>0.$
When $c=0$, (23) reduces to  
\begin{equation}
L\left\{R_{\nu,\mu}(a,0,t)\right\}=\frac{p^\mu}{p^\nu-a}, Re(\nu-\mu)>0, Re(p)>0.
\end{equation}
Finally we recall the definition of Riemann-Liouville  operator of fractional integration in the form
\begin{equation}
_0D_t^{-\nu}f(t)= \frac{1}{\Gamma(\nu)}\int_0^t(t-u)^{\nu-1}f(u)du,
\end{equation}      
with    $_aD_t^0 f(t) = f(t)$ [21,22,23].
The standard kinetic equation, when integrated, yields
\begin{equation}
N_i(t)-N_0=c\;_0D_t^{-1}N(t),
\end{equation}
where $_0D_t^{-1}$  is the standard Riemann-Liouville integral operator. Here it can be mentioned that in the original paper of  Haubold and Mathai [24], the number  density of the species  $i, N_i = N_i(t)$ is a function of time and $N_i(t=0) = N_0$   is the number  density of species  $i$ at time  $t= 0$. If we drop the index $i$ in (27) and replace $c$ by $c^\nu$, then the solution of the generalized equation 
 \begin{equation}
N(t)-N_0=-c^\nu\;_0D_t^{-\nu}N(t),
\end{equation}
is obtained, Haubold and Mathai [24]  as
\begin{equation}
N(t)=N_0\sum^\infty_{k=0}\frac{(-1)^k(ct)^{k\nu}}{\Gamma(k\nu+1)}.
\end{equation}
By virtue of (1), (29) can be written in a compact form as 
\begin{equation}
N(t)  = N_0E_\nu(-c^\nu t^\nu).
\end{equation}                                                             
 In the following, we investigate the  solution of two more  generalized fractional kinetic equations. The results are obtained in a compact form in terms of generalized Mittag-Leffler functions and are suitable for computation. A detailed account of the operators of fractional integration and their applications is available from a recent survey paper of Srivastava and  Saxena [25]. 

\section{Generalized fractional kinetic equations} 
{\bf Theorem 1.} If  $c > 0, \nu>0, \mu>0$, then for the solution of the equation 
\begin{equation}
N(t)-N_0t^{\mu-1}E^\gamma_{\nu, \mu}[-c^\nu t^\mu]=-c^\nu _0D_t^{-\nu}N(t),
\end{equation}   
there holds the formula
\begin{equation}
N(t)=N_0t^{\mu-1}E_{\nu,\mu}^{\gamma+1}(-c^\nu t^\nu).
\end{equation}
{\bf Proof.} By the application of convolution theorem of Laplace transform (Erd\'elyi et al. [26]) we see that  (26) can be written as 
\begin{eqnarray}
L\left\{_0D_t^{-\nu}f(t);p\right\}& = &L\left\{\frac{t^{\nu-1}}{\Gamma(\nu)}\right\}L\left\{f(t)\right\},\\
& = & p^{-\nu}F(p),
\end{eqnarray}                                                                                 
where   $F(p)  = \int^\infty_0 e^{-pu} f(u)du, Re (p)>0$.
Projecting  the equation  (31) to Laplace transform , we obtain 
\begin{eqnarray}
N(t)=L[N(t);p] & = & N_0\frac{p^{-\mu}[1+(p/c)^{-\nu}]^{-\gamma}}{[1+(p/c)^{-\nu}]}\nonumber\\
& = & N_0p^{-\mu}[1+(p/c)^{-\nu}]^{-(\gamma+1)}.
\end{eqnarray}                                            
On using the formula (16), we find that 
\begin{equation}
L^{-1}[p^{-\mu}\left\{1+(p/c)^{-\nu}\right\}^{-(\gamma+1)}]=t^{\mu-1}E^{\gamma+1}_{\nu, \mu}(-c^\nu t^\nu).
\end{equation}      
The result (32) now readily follows from (36).\par 
If we set $\gamma=1$ then (n)! is cancelled. Then in view of the formula         
\begin{equation}
\beta E^2_{\beta, \gamma}(z) = E_{\beta, \gamma-1}(z)+ (1-\gamma+\beta) E_{\beta, \gamma}(z),
\end{equation}
which follows from the definition  of $E^\gamma_{\alpha, \beta}(z)$ given by  (3) ,we arrive at\\[0.3cm]  
{\bf Corollary 1.1.} If $c > 0, \mu>0, \nu>0$, then for the solution of
\begin{equation}
N(t)-N_0 t^{\mu-1} E_{\nu, \mu}[-c^\nu t^\nu] = -c^\nu\; _0D_t^{-\nu}N(t),
\end{equation}      
there holds the formula 
\begin{equation}
N(t) = \frac{N_0t^{\mu-1}}{\nu}[E_{\nu, \mu-1}(-c^\nu t^\nu)+(1-\mu+\nu)E_{\nu, \mu}(-c^\nu t^\nu)].
\end{equation}        
When $\gamma=2$, then by virtue of the following identity
\begin{eqnarray}
E^3_{\beta, \gamma}(z)&=&\frac{1}{2\beta^2}[E_{\beta, \gamma-2}(z)+(3\beta+3-2\gamma)E_{\beta, \gamma-1}(z)\nonumber\\  
&+& \left\{2\beta^2+\gamma^2+3\beta-2\gamma-3\beta\gamma+1\right\}E_{\beta,\gamma}(z)],
\end{eqnarray}
which follows as a consequence of the definition  (3) , we obtain\\[0.3cm]
{\bf Corollary 1.2.}  If $c > 0, \nu>0, \mu>0$, then for the solution of 
\begin{equation}
N(t)-N_0 t^{\mu-1}E^2_{\nu, \mu}[-c^\nu t^\nu]= -c^\nu _0D_t^{-\nu}N(t),
\end{equation}        
there holds the relation 
\begin{eqnarray}
N(t)& = & N_0 t^{\mu-1}E^3_{\nu,\mu}(-c^\nu t^\nu),\\      
 & = & \frac{N_0 t^{\mu-1}}{2\nu^2}\left[E_{\nu,\mu-2}(-c^\nu t^\nu)+\left\{3(\nu+1)-2\mu\right\}E_{\nu, \mu-1}(-c^\nu t^\nu)\right.\nonumber\\                     
& + & \left.\left\{2(\nu^2+\mu^2)+3\nu-2\mu-3\nu\mu+1\right\}E_{\nu,\mu}(-c^\nu t^\nu)\right].
\end{eqnarray}
Next, if we set $\gamma=0$, then by virtue of the identity $E^0_{\nu,\mu}(z)=\frac{1}{\Gamma(\mu)}$, we arrive at another result given by Saxena et al. [1].\\[0.3cm]
{\bf Theorem 2.} If  $c > 0 , b\geq 0, Re (p ) > 0 , \nu > \mu+1$, then for the solution of 
\begin{equation}
N(t)-N_0R_{\nu, \mu}(-c^\nu, b,t)=-c^\nu\; _0D_t^{-\nu}N(t),
\end{equation}    
there holds the formula 
\begin{equation}
N(t)=\frac{N_0}{\nu}(t-b)^{\nu-\mu-1}[E_{\nu,\nu-\mu-1}(-c^\nu(t-b)^\nu)+(\mu+1)E_{\nu,\nu-\mu}(-c^\nu(t-b)^\nu)].
\end{equation}
{\bf Proof.} Taking Laplace transform of  both sides of (45), it gives
\begin{eqnarray*}
N(t)& = & L\left\{N(t);p\right\}=N_0 L^{-1}\left[\frac{e^{-bp}p^{\mu-\nu}}{\left\{1+(c^\nu/p^\nu)\right\}^2}\right]\\
& = & N_0 L^{-1}\left[e^{-bp}p^{\mu-\nu}\sum^\infty_{n=0}\frac{(2)_n}{(n)!}(\frac{-c^\nu}{p^\nu})^n\right]\\
& = & N_0\sum^\infty_{n=0}\frac{(2)_n(-c)^{n\nu}}{(n)!}L^{-1}[e^{-bp}p^{\mu-\nu-n\nu}]\\
& = & N_0(t-b)^{\nu-\mu-1}\sum^\infty_{n=0}\frac{(2)_n(-c)^{n\nu}(t-b)^{n\nu}}{(n)!\Gamma(n\nu+\nu-\mu)}\\
& = & \frac{N_0}{\nu}(t-b)^{\nu-\mu-1}\sum^\infty_{n=0}\frac{(-c)^{n\nu}[(n\nu+\nu-\mu-1)+(1+\mu)](t-b)^{n\nu}}{\Gamma(n\nu+\nu-\mu)}\\
& = & \frac{N_0}{\nu}(t-b)^{\nu-\mu-1}\left[\sum^\infty_{n=0}\frac{\left\{-c^\nu(t-b)^\nu\right\}^n}{\Gamma(n\nu+\nu-\mu-1)}+(\mu+1)\sum^\infty_{n=0}\frac{\left\{-c^\nu(t-b)^\nu\right\}^n}{\Gamma(n\nu+\nu-\mu)}\right]\\
& = & \frac{N_0}{\nu}(t-b)^{\nu-\mu-1}[E_{\nu,\nu-\mu-1}\left\{-c^\nu(t-b)^\nu\right\}+(\mu+1)E_{\nu,\nu-\mu}\left\{-c^\nu(t-b)^\nu\right\}],
\end{eqnarray*}  
which is same as (45). This completes the proof of  theorem 2.
If we set $\mu=0$, theorem 2 reduces to\\[0.3cm] 
{\bf Corollary 2.1.} If $c> 0 , b\geq 0, \nu>1$, then for the solution of
\begin{equation}
N(t)-N_0R_{\nu,0}(-c^\nu, b,t)=-c^\nu\;_0D_t^{-\nu}N(t),
\end{equation}
there holds the formula
\begin{equation}
N(t) = \frac{N_0}{\nu}(t-b)^{\nu-1}[E_{\nu,\nu-1}(-c^\nu(t-b)^\nu)+E_{\nu,\nu}(-c^\nu(t-b)^\nu)].
\end{equation}
For $b = 0$, theorem 2 yields\\[0.3cm]
{\bf Corollary 2.2.} If $c > 0, \nu>\mu+1$, then for the solution of 
\begin{equation}
N(t)-N_0R_{\nu,\mu}(-c^\nu,0,t)=-c^\nu\; _0D_t^{-\nu}N(t),
\end{equation}
there holds the formula  
\begin{equation}
N(t)=\frac{N_0}{\nu}t^{\nu-\mu-1}[E_{\nu,\nu-\mu-1}(-c^\nu t^\nu)+(\mu+1)E_{\nu, \nu -\mu}(-c^\nu t^\nu)].
\end{equation}
If we further take $\mu=0$ then the above corollary reduces to the following result:\par
        If $c > 0 , \nu>1$, then the solution of 
\begin{equation}
N(t)-N_0 F_\nu[-c^\nu,t]=-c^\nu D_t^{-\nu}N(t),
\end{equation}       
is given by 
\begin{equation}
N(t)=\frac{N_0t^{\nu-1}}{\nu}[E_{\nu,\nu-1}(-c^\nu t^\nu)+E_{\nu,\nu}(-c^\nu t^\nu)].
\end{equation}\par       
\bigskip
\noindent
{\bf References}\par
\medskip
\noindent
[1] R.K. Saxena, A.M.Mathai, H.J. Haubold, Astrophys. and \par
Space Sci., 282 (2002) 281. \par
\noindent
[2] J. Jorgenson, S. Lang, in: Mathematics Unlimited - 2001\par
and Beyond, Eds. B. Engquist, W. Schmid, Springer-Verlag,\par 
Berlin and Heidelberg 2001, 655.\par
\noindent
[3] R. Hilfer, in : Applications of Fractional Calculus in Physics,\par 
Ed. R. Hilfer, World Scientific, Singapore 2000, 1.\par
\noindent
[4] C. Tsallis, in: Nonextensive Entropy: Interdisciplinary Applications,\par 
Eds. M. Gell-Mann, C. Tsallis, Oxford University Press,\par 
New York 2003,1.\par
\noindent
[5] G.M. Mittag-Leffler, C.R. Acad. Sci. Paris 
                    (Ser.II )137 (1903) 554.\par
\noindent 
[6] G.M. Mittag-Leffler, Acta Math. 29 (1905) 101.\par
\noindent
[7] T.R. Prabhakar,  Yokohama Math. J. 19 (1971) 7.\par
\noindent
[8] A. Wiman, Acta Math. 29 (1905) 191.\par
\noindent
[9] A. Wiman,  Acta Math. 29 (1905) 217.\par
\noindent
[10] R.P. Agarwal, C.R. Acad. Sci. Paris  236 (1953) 2031.\par
\noindent
[11] P. Humbert, C.R. Acad. Sci. Paris  236 (1953) 1467.\par
\noindent
[12] P. Humbert, R.P. Agarwal, Bull. Sci. Math. (Ser.II) 77 (1953)
180.\par
\noindent
[13] A. Erd\'elyi, W. Magnus, F. Oberhettinger, F.G. Tricomi,  Higher\par
Transcendental Functions, Vol. 1, McGraw-Hill, New York, Toronto\par
and London 1953.\par
\noindent
[14] A.M. Mathai, R.K. Saxena,  The H-function with Applications\par 
in Statistics and Other Disciplines, Halsted Press, John Wiley \& Sons,\par 
New York-London-Sydney-Toronto 1978.\par
\noindent
[15] E.M. Wright, J. London Math. Soc. 10 (1935) 286.\par
\noindent
[16] E.M. Wright, Proc. London Math. Soc.  46 (1940) 389.\par
\noindent
[17] T.T. Hartley, C.F. Lorenzo, NASA 1999/ TP-1998-208693 (1998) 1.\par
\noindent
[18]  Y.N. Robotnov, Tables of a Fractional Exponential Function of \par
Negative   Parameters and its Integral (in Russian), Nauka, Russia 1969.\par
\noindent
[19] Y.N. Robotnov,  Elements of Hereditary Solid Mechanics\par 
(in English), MIR Publishers, Moscow 1980.\par
\noindent
[20] C.F. Lorenzo, T.T. Hartley,  NASA / TP-1999-209424 (1999) 1.\par
\noindent
[21] K.B. Oldham, J. Spanier, The Fractional Calculus: Theory\par 
and Applications of Differentiation and Integration to Arbitrary Order,\par 
Academic Press, New York 1974.\par
\noindent
[22] S.G. Samko, A.A. Kilbas, O.I. Marichev,  Fractional Integrals\par 
and Derivatives. Theory and Applications, Gordon and Breach,\par 
Sci. Publ., New York 1990.\par
\noindent
[23] K.S. Miller, B. Ross, An Introduction to the Fractional\par 
Calculus and Fractional Differential Equations, John Wiley \& Sons,\par
 New York 1993.\par
\noindent
[24] H.J. Haubold, A.M. Mathai,  Astrophys. Space Sci.\par 
327 (2000) 53.\par
\noindent
[25] H.M. Srivastava, R.K. Saxena, Appl. Math. Comput.\par 
118 (2001) 1.\par
\noindent
[26] A. Erd\'elyi, W. Magnus, F. Oberhettinger, F.G. Tricomi, \par 
Tables of Integral Transforms, Vol. 1, McGraw-Hill, New York, Toronto\par 
and London 1954.\par
\noindent
[27] R. Metzler, J. Klafter, Phys. Rep. 339 (2000) 1.
\end{document}